# Inline Visualization and Manipulation of Real-Time Hardware Log for Supporting Debugging of Embedded Programs

ANDREA BIANCHI, Industrial Design & School of Computing, KAIST, Republic of Korea
ZHI LIN YAP*, School of Computing, KAIST, Republic of Korea
PUNN LERTJATURAPHAT*, Industrial Design, KAIST, Republic of Korea
AUSTIN Z. HENLEY, Microsoft, USA
KONGPYUNG (JUSTIN) MOON, Industrial Design, KAIST, Republic of Korea
YOONJI KIM, College of Art & Technology, Chung-Ang University, Republic of Korea

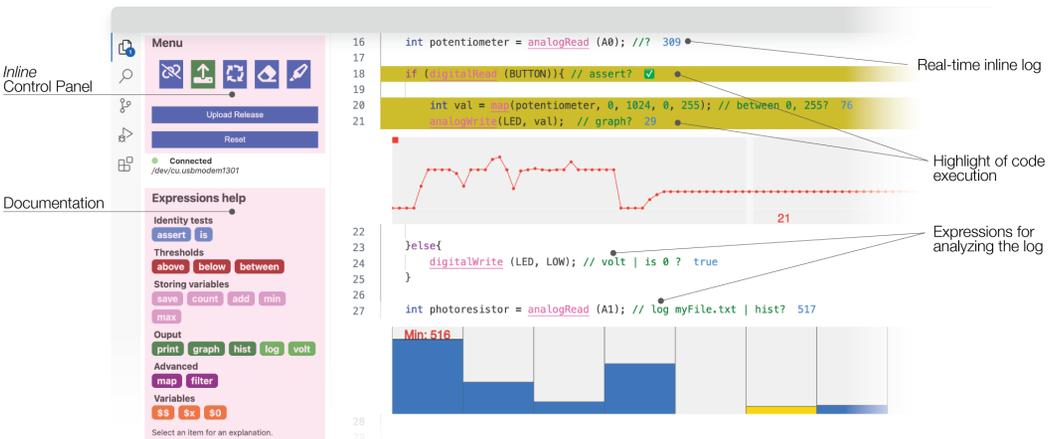

Fig. 1. INLINE is a tool that augments the code editor by visualizing the real-time state generated on an embedded hardware platform. Logs are displayed directly within the code as inline text annotations, glyphs, and graphs, and a line highlighter to indicate which line is currently executed. The user can filter and manipulate the output logs using a rich expression language embedded in the comments placed next to an instrumented function call.

The advent of user-friendly embedded prototyping systems, exemplified by platforms like Arduino, has significantly democratized the creation of interactive devices that combine software programs with electronic hardware. This interconnection between hardware and software, however, makes the identification of bugs

---

*These authors contributed equally to this work

Authors' Contact Information: Andrea Bianchi, Industrial Design & School of Computing, KAIST, Republic of Korea, andrea@kaist.ac.kr; Zhi Lin Yap*, School of Computing, KAIST, Republic of Korea, zhilin@kaist.ac.kr; Punn Lertjaturaphat*, Industrial Design, KAIST, Republic of Korea, punnlert@kaist.ac.kr; Austin Z. Henley, Microsoft, USA, azh321@gmail.com; Kongpyung (Justin) Moon, Industrial Design, KAIST, Republic of Korea, jkpmoon@kaist.ac.kr; Yoonji Kim, College of Art & Technology, Chung-Ang University, Republic of Korea, yoonji.h.kim@gmail.com.

---

very difficult, as problems could be rooted in the program, in the circuit, or at their intersection. While there are tools to assist in identifying and resolving bugs, they typically require hardware instrumentation or visualizing logs in serial monitors. Based on the findings of a formative study, we designed INLINE a programming tool that simplifies debugging of embedded systems by making explicit the internal state of the hardware and the program's execution flow using *visualizations of the hardware logs directly within the user's code*. The system's key characteristics are 1) an inline presentation of logs within the code, 2) real-time tracking of the execution flow, and 3) an expression language to manipulate and filter the logs. The paper presents the detailed implementation of the system and a study with twelve users, which demonstrates what features were adopted and how they were leveraged to complete debugging tasks.

## 1 INTRODUCTION

The advent of user-friendly embedded prototyping systems, exemplified by platforms like Arduino[1], has significantly democratized the creation of interactive devices that combine software programs with electronics, sensors, and actuators—a domain commonly referred to as *physical computing* [33]. Because of the popularity of physical computing among non-technical users and children [37], prototyping with embedded systems is often mistakenly perceived as a simple activity ideal for introductory STEM courses, when in fact, it requires the understanding of both programming and electronics and of how these unfold together. Previous work, for example, reported that 54% of the unsolved problems in a typical physical computing exercise emerged from the interplay between the circuit and the software, with hardware errors (e.g., circuit assembly or wrong components) often misrecognized as program bugs [2]. This interconnection between hardware and software makes debugging time-consuming and identifying the root cause of errors and bugs challenging, causing confusion and frustration.

Researchers within the software development community provide an explanation for this misdiagnosis of the true causes of an error. Understanding a program's state involves deducing and interpreting the program's internal state from externally observable information (i.e., *information barriers* [21]). Unlike conventional end-user programming, which offers an array of options for tracing the internal state of the code (symbolic debugging environments [20, 26] and advanced logging capabilities [14]), embedded software systems typically require manual instrumentation of the circuit with additional hardware. For example, on-chip debuggers via JTAG, and tools for hardware inspections [9, 31, 46] facilitate identifying errors but require hardware expertise or specialized equipment beyond the skill set of the average maker. Alternatively, it is possible to rely on physically observable proxies, such as actuators and blinking LEDs used to externalize the inner hardware state, but this approach does not scale and it is limited to simple debugging cases.

For these reasons, print statements are the most ubiquitous form of debugging and the preferred method by beginners [14]. Print statements produce a real-time textual output that is visible on terminal monitors, and do not require additional hardware. However, identifying and tracing asynchronous logs invoked in different parts of the code is ineffective, as these appear all together in the same terminal window and in rapid succession. While enhanced logging tools exist for end-user software development [14], these do not work with embedded systems. Overall, the problem with external representations like physical proxies and printed logs is that they are often **hard to interpret or to map to the code** that generated them [14]. Furthermore, they are also **difficult to trace in real-time**, offering *limited visibility* [31] of the internal state of the embedded code.

---

[1]https://www.arduino.cc

Inspired by Bret Victor's notion of *representations of dynamic behavior* [47], this paper aims to address this problem by making explicit the link between the program's code and the inner state of the hardware that results from the code execution. We propose Inline, a software tool integrated into a traditional code editor that simplifies the debugging of embedded systems (e.g., an unmodified Arduino) by *logging and visualizing* the internal state of the hardware (e.g., pin state, results from function calls, voltage levels, execution flow...) directly within the program's code—*inline* with the code instruction that triggered the hardware change. These visualizations are updated in real-time using text annotations, line highlightings, glyphs, or graphs that are displayed within the existing and familiar programming environment of a code editor rather than external tools [31] and can be stored in logs for further offline analysis. Furthermore, using a simple expression language written as a mark-up superset of the program's comments (e.g., similarly to how JSDoc is used to annotate JavaScript), developers can manipulate and filter the real-time values from the hardware, allowing them to better inspect the code and determine the causes of bugs.

## 1.1 A practical example

To give an example, Figure 1 shows the code for reading the value from a potentiometer using the built-in Analog-to-Digital Converter/ADC (i.e, `analogRead` at line 16) and mapping it (i.e, `map` at line 20) to the brightness of an LED by changing the duty-cycle of a pin driven via Pulse-Width Modulation (i.e., `analogWrite` at line 21) only if a button is pressed (i.e., `digitalRead` at line 18). Although this is a simple program, many things could go wrong. The potentiometer could be incorrectly wired to the wrong pin, or the button could be connected without an appropriate pull-down resistor, and finally, the 10-bit value from the ADC could be incorrectly mapped to a suitable 8-bit value for the duty cycle. More importantly, these functions return and pass values that remain *invisible* within the program flow, unless the developer chooses to deliberately insert print statements to generate console logs. Not only these additional statements are cumbersome as they have to be manually added/removed, but they could also be a source of bugs themselves.

Instead, with Inline the developer interested in reading the value from the ADC can simply place the comment //? inline with the appropriate instruction and see the real-time value from the hardware without using any additional print statement (*309* at line 16). Using expressions within the comments, the developer can further specify conditions of interest or formatting properties, for example indicating an interest only for log values within a certain threshold (//between 0, 255? at line 20) or querying a single pin for its digital state with an assertion (//assert? at line 18, resulting in the  symbol for digital HIGH). Alternatively, the developer can opt for visualizing the log using voltage (//volt ? at line 24) or using a plot directly juxtaposed with the code. The incoming stream of logs can be graphed with a line (//graph? at line 21) or using a histogram (//hist? at line 27). Finally, the operations can also be concatenated using the *pipe operator* "|", allowing, for example, to store the logs in a textual record while at the same time graphing them (line 27). These are just some of the unique possibilities offered by Inline.

## 1.2 Contributions

In the rest of this paper, we present a survey of the literature, followed by a formative study with 10 participants that was used to gather requirements and inspire the features of the system. We then present INLINE and its working principles, as well as a full description of the proposed expression language. We finally present the benefits of our approach via a user study with 12 participants.

Through our studies and system design we contribute:

(1) An account of the typical usage of debugging logs for physical computing, and a description of the opportunities emerging from better linking hardware to software. We present a list of challenges and how these map to design goals and system features.

(2) We present the INLINE system that supports debugging by providing better visualization and logging capabilities. INLINE allows visualizing and manipulating the real-time hardware data and state directly within the code editor, using a simple expression language. This real-time behavior is conceptually close to the notion of *live programming* [38], as INLINE provides an *immediate* visual feedback (i.e., time and line location) describing the dynamic behavior of a program (i.e., the hardware log) while programming.

(3) A characterization of how users adopt INLINE and use its features to explore data to identify better errors or bugs spread between the program software and the circuit hardware.

## 2 RELATED WORK

The following section introduces tools for debugging in the domains of software engineering and physical computing. While physical computing draws numerous inspiration from software engineering practice and coding styles [7, 27] and test-driven development [16, 31, 49], it still lacks the rich ecosystem of tools, editors, libraries, and debuggers typical of end-user programming [2]. The following section highlights the common problems of the two fields and the different approaches to solving them, with emphasis on debugging and logging tools.

### 2.1 Tools for Debugging Software

Understanding a program's execution flow requires inferring and interpreting the *program's internal state* from externally observable information [21], or, as simply put by Brad Fitzpatrick in an interview from *Coders at Work*[41], "trying to understand the root cause of things". For example, common questions programmers raise include finding the source of a variable's value in the code or understanding why a function is or is not being called [23, 36, 42]. Towards fulfilling these information needs, laboratory studies found that 35–50% of programmers' time is spent navigating code and program output [22, 36], and programmers are often incorrect about where they believe the information is [35]. These findings motivated numerous work and research for improving coding tools and practices for end-user programming, including a plethora of features for integrated development environments (IDEs) or AI-capable tools to assist programmers during debugging.

Standard tools to debug software include logging information to the console or a file and using a debugger to step through line by line or to halt at a breakpoint. More sophisticated approaches include augmenting the code editor with live log information while the program is executing, such as QuokkaJS [2] and Console Ninja [3], which both display the result of expressions in the editor. Log-it drastically improves working with output logs by providing an interactive and visual interface that enables programmers to make sense of log information as they are debugging [14].

Another approach taken by debugging tools involves recording information about the program execution. For example, Theseus addresses misconceptions about program behavior by displaying how many times each function has been executed [26]. Whyline enables programmers to ask *why* and *why not* questions about the program behavior, though it comes at substantial performance costs even for simple programs [19, 20]. For debugging web apps, Timelapse allows programmers to replay program executions and pause at specific points to determine the cause of bugs [4]. Another tool, CodeDeviant, records the inputs and outputs of code modules at every program execution so that it can warn the programmer when a function's behavior has changed [13].

---

[2]https://quokkajs.com
[3]https://console-ninja.com

More recently, AI-driven code generation tools built on top of *large-language models* (LLMs) have gained significant attention in the field of software development. Prominent examples of such tools include chatGPT and Github Copilot. These AI-powered systems exhibit the potential to assist developers by generating code based on natural language prompts. However, the quality of the generated code often depends on the clarity of the input prompts and the context provided [15]. Several AI-powered tools for debugging 3D [30] and 2D [11] games, spreadsheets [40], or that provide code explanations [28] have been experimented with, but these tools raise concerns regarding proper learning, over-reliance and plagiarism [15].

The INLINE system presented in this paper draws numerous inspirations from the above work, but differently from all these, it deals with the *domain of physical computing and embedded systems*, where debugging requires inspecting and understanding the state of *both* the software program and of the electronic hardware.

## 2.2 Tools for Debugging Hardware

Debugging embedded systems further exacerbates the problems of programmers satisfying information needs—opaque state information, variability from sensors or timing, and limited debugging tools. This section delves into the diverse array of tools and techniques available to facilitate efficient hardware debugging.

A critical aspect of hardware development revolves around the construction of circuits. To aid this process, tools like HeyTeddy [16] have emerged to streamline the coding process, catering especially to beginners. HeyTeddy offers a conversation-based prototyping experience that not only aids in circuit assembly but also provides assistance in testing electronic components like sensors and motors. Complementing these efforts are other notable works aimed at simplifying circuit construction. These include solutions that reduce the complexity of wiring [5, 8, 39] and others that eliminate the need for physical wires altogether [24, 48]. Users can thus focus on the functionalities of electronic components like actuators and sensors without the added complexity of physical wiring. Furthermore, Flowboard [3], a device that combines an iPad with an Arduino and a switchboard circuit beneath it, along with breadboards on each side, facilitates the embedding of interaction into physical circuits. Users can easily connect these processing nodes to the breadboards, creating interactions between sensors and actuators. Finally, other words, focused on practices and software systems for designing and understanding circuits [1, 7, 18, 27, 29], and methods to blend hardware components with digital circuits [51].

Effective hardware debugging necessitates either external instruments like oscilloscopes and digital multimeters or instrumenting a circuit with testing hardware (e.g., JTAGs or active breadboards). Yet, conventional methods, such as using oscilloscopes and digital multimeters, typically entail laborious modifications to the board's connections. Alternatively, several techniques for voltage monitoring [9, 32] and current visualization [52] offer non-invasive approaches that obviate the need for connection modifications. However, these methods require the construction of special devices, often in the shape of instrumented breadboards. Several approaches even combined hardware debugging with augmented reality. VirtualComponent [17] empowers users to place virtual electronic components on physical breadboards, simplifying value adjustments and providing live observation of circuit changes. Meanwhile, ARDW [6] enhances the debugging process by minimizing the need to switch between PCB and design files. However, it's crucial to acknowledge that ARDW's setup complexity entails the use of equipment such as a multimeter, overhead projector, and probes.

Integration of hardware inspection with software debugging is another crucial aspect of the debugging process. CircuitSense [53] automatically generates real-time virtual representations of circuits. However, it has to be integrated with other software tools, like ToastBoard [9], to enable effective debugging. Simpoint [44], Pinpoint [46], Scanalog [43], and Bifröst [31] all offer real-time hardware signal visualization, but with distinct features and capabilities, ranging from comparing real-time hardware signals with simulated models to programmable signal probing and variable value inspection. These tools provide information about the execution of a program (e.g., individual signal values by hovering with the mouse on a graph, as in Bifröst) but the visualizations and the interactions with the hardware data happen outside the direct context of the program's code using separate graphical interfaces.

Like these systems, we aim to enhance the debugging process by bridging the gap between code and hardware. However, differently from past works, INLINE provides augmented logging capabilities to traditional hardware leveraging real-time visualizations of the embedded system's inner state *directly within the code editor*, and supporting *a set of expressions for filtering and manipulating* the streams of hardware logs.

## 3 FORMATIVE STUDY

To understand the challenges that non-technical users face when developing and debugging interactive prototypes using embedded systems (e.g., Arduino), we recruited 10 participants: 8 makers—7 male, 1 female, average aged 28.3 (SD:2.3, Min: 26 and Max:31) and 2 university professors (both male, 2-15 yrs of teaching experience). For details about prior experiences, please refer to Table 1.

Table 1. Formative study participants' demographics and expertise.

| Participants | Occupation | Department | Age | Physical Computing Experience (years) | Software Development Experience (years) | Physical Computing Self-Rank (scale:1-7) |
|---|---|---|---|---|---|---|
| P1 | maker | Design (M.S) | 26 | 9 | 6 | 6 |
| P2 | maker | Design (Ph.D) | 28 | 7 | 7 | 5 |
| P3 | maker | Design (M.S) | 26 | 6 | 6 | 6 |
| P4 | maker | CS (Ph.D) | 26 | 2 | 8 | 3 |
| P5 | maker | Design (Ph.D) | 29 | 10 | 7 | 6 |
| P6 | maker | CS (Ph.D) | 31 | 7 | 7 | 6 |
| P7 | maker | CS (Ph.D) | 32 | 6 | 14 | 4 |
| P8 | maker | Design (Ph.D) | 28 | 8 | 6 | 6 |
| P9 | instructor | Electronics | 50 | 15 (physical computing) + 15 (teaching) | - | - |
| P10 | instructor | Design | 35 | 12 (physical computing) + 2 (teaching) | - | - |
| Avg (SD) | - | - | 28.3 (2.3) | 9.9 (7.7) | 7.6 (2.7) | 5.3 (1.2) |

We conducted semi-structured interviews, each lasting about 50 minutes, to investigate challenges makers encountered in their past prototyping experiences. We initially asked the participants to recall a project they worked on and to describe it. We then asked them to highlight their typical workflow, difficulties, and their debugging strategies. Following the transcription and translation to English of these audio interviews, we applied open and axial coding for analysis. Below, we outline three key challenges we identified across the participants that occur during development and debugging. Overall, our observations pinpointed three primary obstacles in developing and debugging with hardware logs: 1) navigating numerous text logs on the serial monitor, 2) managing and comprehending the flow of the code execution, and 3) inspecting and interpreting log data.

## 3.1 Challenge 1: Printing and Managing Logs on a Serial Monitor

Printing input and output values on a serial monitor is a common practice among programmers. These textual logs help makers understand what is happening behind the code (P2, P5, and P6), and for embedded systems, the logs are typically displayed as a stream of line-separated entries in serial monitor windows (e.g., the Arduino Serial Monitor). Instructors stressed that they explicitly teach students how to use the serial monitor for printing logs, "otherwise, students have no idea what's going on in their Arduino (P10)."

The main difficulty is, however, when users have to deal with **multiple print statements** that produce logs in seemingly scattered order and that cannot be immediately traced back to their sources. For instance, P7 said: "it was almost impossible to follow all different data in the text logs. [...] Even when I wanted to trace them, the scroll bar kept moving on, and I lost them." A possible solution explored by the participants is to maintain a structure for the log entries using dividers, commas, labels, and even timestamps, so to ensure a more legible and understandable output (P2-P4, P7, and P8). However, multiple participants remarked that these strategies are not always effective. P6 explained: "I put too many text logs between [instructions], so I get confused later on. Even if I try to organize them, it doesn't work out, and they get mixed up like that again later [i.e., when displayed on the terminal window]." As a final result, some participants opted for logging their data to files and exploring them later on, trading off real-time logs for output that was easier to interpret: "I had to record the data rather than monitor it in real-time due to the complexity of numerous variables involved, even when accessing live feeds was essential (P2)".

A workaround to having multiple logs is printing only a few at a time. However, this strategy requires modifying the code each time. P8 explained that removing printouts might result in bugs and a waste of time: "If you add and remove annotations like this, you have to recompile every time. I think that's really time-consuming. If someone wants to view only specific logs, it means recompiling everything every time there's a change." Alternatively, P4 suggested that it would be nice if it was possible to selectively toggle logs without compiling the code, saying that "it would be really nice if I could simply toggle the output of the logs I want to see in real-time, on and off."

## 3.2 Challenge 2: Tracking Code Execution In Real-Time

Primarily users are interested in **real-time behavior** of their devices and use logs while interacting with the device, stressing the importance of *live prototyping*. P3, recalling an interactive project that leverages an accelerometer for sensing motion, stated that when "I [move my prototype], then I get to see how the [sensor] values change immediately." Similarly, P4 stated that he "had to look at the hardware logs in real-time, [because] I wanted to double-check that the changes I made were actually taking effect immediately while I was adjusting things right there on the spot."

Real-time logs also help determine which part of the code is called, allowing users to track the execution flow or spot unreachable parts of the code. However, mapping real-time logs back to the code is challenging, as the code is not executed in the same order as it is written in the text. In the words of P6, "the logs were printed from functions lower down, but I needed to make changes in a middle function [...] I had one window displaying the code and another for the logs [...] I needed to scroll back and forth in both windows to locate the code that needed fixing and keep track of the logs, which confused me."

An alternative to this approach is to dismiss altogether print statements and instead rely on the behavior of physical proxies (e.g., LED and actuators) to determine whether a part of the code has been reached, or a certain condition has been satisfied. This approach is cumbersome but provides an immediate and real-time interpretation of what is happening. As P1 said, "it demands much less cognitive effort, resulting in quicker reaction time because of the simple LED states, either on or off. For instance, if you expected it to be off but unexpectedly turned on, you could recognize that something was wrong. [...] Identify the exact location and then, I guess, fix it with just debugging afterward."

### 3.3 Challenge 3: Making sense of the logs

While managing and tracking logs is challenging, **filtering and making sense of their data is also a complex activity**, for which there is no clear solution. All participants (makers and instructors) stressed the difficulties of analyzing data expressed in the text logs. As P9 stated, "students try to print and read text logs to debug their prototypes, but they cannot explain the meaning of those texts even though they wrote the code." P10, the other instructors, explained that students struggle in filtering the appropriate data they need: "I feel like [students] don't do well at debugging at all. They might look at this data, but it's not helpful for them."

One method our participants consistently used to filter data was by employing thresholds and cutoff conditions. For example, P3 printed the logs to "only outputting data when it goes outside that range or only when it falls within that range." However, determining these ranges requires both expertise and additional code structures (i.e., branching structures, as *if-statements*). In some cases, the conditions introduced by the participants to debug a problem were indeed the cause of new bugs. For example, P2 said that "I adjusted the cutoff value to make the data smoother, then it started to be too slow and did not match the real-time movement [...] So, I increased the beta value until it matched the speed, and then I started to have noisy data. [...] Those two variables interacted, so I spent much time doing trials and errors to keep adjusting the values to the proper amount."

### 3.4 Summary of findings

Overall, this section shows that print statements are ubiquitous not only in software development [14] but also for embedded programming, and it highlights the problems of dealing with streams of logs from multiple sources and the requirements to recompile the code each time the logs have to be updated. Participants also expressed the importance of being able to debug events in real-time directly when sensors are reading the data or certain parts of the code are executed. The liveliness aspect of the log is at times the most important feature, so users are willing to trade off detailed logs for rapidly identifiable signals (e.g., using an LED that lights up). Finally, making sense of the logs is also a challenging activity, and for this reason, it is often limited to setting boundary conditions and thresholds. When introducing control structures to support these strategies, new bugs often emerge (as also seen in previous work [2]), ultimately making debugging even harder.

## 4 THE INLINE SYSTEM

Inline is a software tool that makes debugging easier by *visualizing and manipulating* the live hardware logs right in the code editor. It works with embedded systems like Arduino, letting you work with the program's code and logs at the same time. Inline operates by instrumenting the user's code with print statements and meta-data, such as line numbers and unique identifiers for expressions, directly within the user's code, generating real-time logs of the hardware state and displaying the results next to the corresponding sections of the code that executes them. This also allows filtering the logs and modifying their appearance/values programmatically without compiling or uploading the code, using a simple expression language embedded within the code comments.

The final design of our system was motivated by the challenges described in the formative study, which we used to inform three **D**esign **G**oals: **DG1**) inline visualization of debugging logs along with code using text, glyphs, and graphs; **DG2**) real-time tracking of the program's execution flow using line highlights; and **DG3**) filters and manipulations of logs using textual expressions embedded in code comments. We describe these goals and their features in detail in the following sections, using a walk-through example to illustrate their practical applicability and benefits. We then describe the implementation of different parts of the system (Section 5) and how code instrumentation was achieved (Section 5.1).

## 4.1 Walk-through: building a music sequencer with Inline

To explain the features and advantages of Inline, we show how it facilitates prototyping a glove music sequencer (see Fig. 3-left) mounting four flex sensors. Notes (i.e., frequencies) are mapped to fingers, while the volume is controlled by how much fingers are bent. Notes are played sequentially in a loop at fixed intervals (beat-per-minute, or BPM) set with a rotary encoder. The sound is produced with a piezo buzzer, triggered only if the user presses a *play* button to prevent accidental noise. The electronics are spread across the glove, and a breadboard with a custom circuit interfaced to an Arduino UNO R3[4] connected via a serial link to a computer.

> ***Walk-through:*** *the user connects each of the four flex sensors (SZH-SEN02) to a resistor to form a voltage divider, with the center tap wired to an analog pin. To read from the Analog-to-Digital Converter (ADC), the user calls the function* analogRead(pin). *The push button and the incremental linear encoder (Alps Alpine EC12E2420803) are directly connected to digital pins using the Arduino internal pull-up resistors using the code* pinMode(pin, INPUT_PULLUP). *The buzzer is connected to a PWM-enabled pin. On the PC, traditionally, a user would attempt to print the values for all the four sensors, perhaps using comma-separated values to list of the readings on a single line (see right of Figure 3). Instead, here the user opens a new Arduino sketch (i.e., a C/C++ file with a .ino extension) using a code editor with Inline installed. Through shortcuts or a side panel, the user connects serially to the Arduino, choosing the baud rate (e.g., 115200) from a list, and the editor confirms the operation's success. Inline parses the file and highlights with*

---

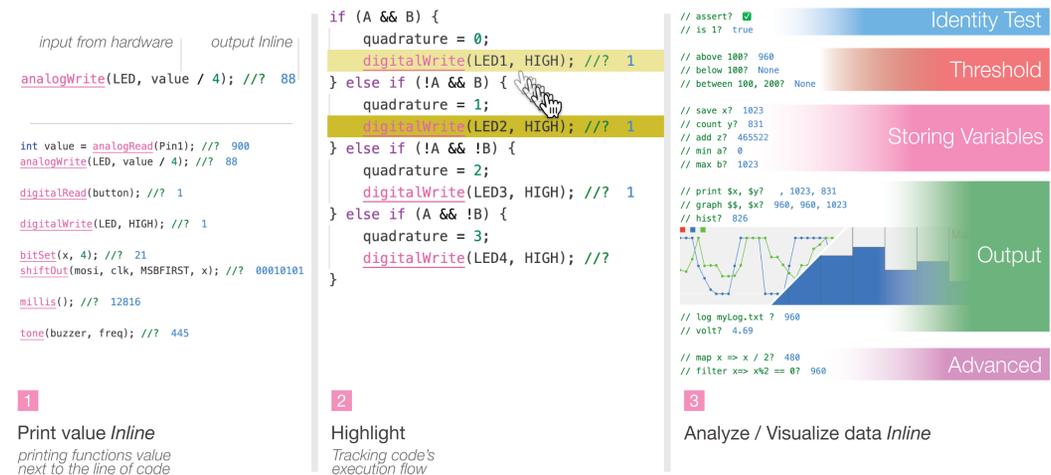

**Fig. 2.** The three **D**esign **G**oals and the specific features of Inline, which were informed from the Challenges that emerged in the Formative Study.

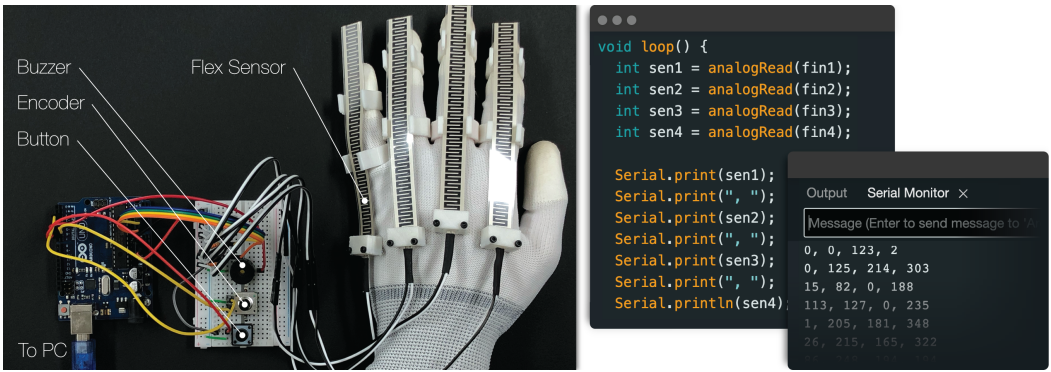

Fig. 3. The music glove prototype and its parts used in the walk-through (left). The flex sensors are connected to analog pins, and their values are printed as comma-separated values in the Arduino IDE serial monitor.

> *an underscore all the native Arduino functions[5] in the file, helping the user identify the instrumented lines (e.g., Figure 8).*

## 4.2 Design Goal 1: Visualizing debugging logs together with code using text, glyphs, or graphics

Debugging logs help understand "what is happening inside the hardware of an embedded system" when a program executes. INLINE displays the logs alongside the code, highlighting the link between the hardware state and the instruction that caused it, without using explicit print statements. Each log is treated as an independent stream of values, making it possible to simultaneously visualize multiple logs that do not affect each other. This approach stands in sharp contrast to the traditional method of accumulating print statements in a console window and then manually tracing them back to the code that generated them, as illustrated on the right side of Figure 3.

INLINE log streams can be visualized using textual annotations, glyphs, and graphs. Textual or numerical annotations are triggered by placing the comment //? at the end of a line (Figure 8-[1]). This prompts the logs to be displayed as live text that updates when new incoming logs for the line are received. Removing the //? disables the log. If no logs are detected on the line, three dots **...** indicate that no value was received.

Alternative visualizations are triggered using *expressions*. The comment //assert? generates the glyph ✅ and ❌ that immediately convey whether the resulting log contains a nullish value—a 0, an empty string, or false—or not (Figure 5). This visualization was inspired by the usage of physical proxies (e.g., LEDs) described in the formative study. Graphical and interactive visualizations are possible with the expressions //graph? and //hist? which append the logs to live graphs and histograms within the code (Figure 4). Line graphs may show multiple signals, support explorations via zoom and pause, and provide details about single data points with mouse hovering. Histograms offer an immediate understanding of the occurrences of a value in user-defined ranges, which can be customized via mouse scrolling. Both graphs implement vertical auto-scaling as the Arduino Serial Plotter[6].

> *Back to the walk-through, the user first checks the readings from the flex sensors using the expression //?. Eyeballing the numerical values displayed inline with each* `analogRead` *function call, the user identifies an unusually high value for one of the sensors. The*

---

[5]https://www.arduino.cc/reference/en
[6]https://docs.arduino.cc/software/ide-v2/tutorials/ide-v2-serial-plotter

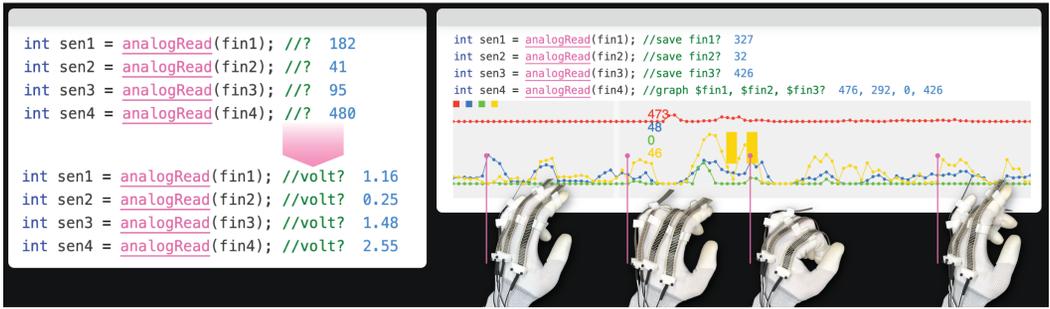

Fig. 4. The user inspects the readings from the flex sensor using the textual annotations //? and //volt? and finds irregular values from the fourth one. Graphs with the combined readings show changes in values over time when the user moves the fingers.

*expression //volt? confirms that the voltage from the fourth flex sensor is much higher than those of other fingers (Figure 4-left). Further inspection via graphs allows us to compare all values over time when the fingers move—sensor 4 indeed operates outside the expected ranges (Figure 4-right). The problem is an incorrectly chosen resistor for the voltage divider, one that's much too large—a common error also reported in previous studies [2]. After replacing it, the readings from the sensor become consistent with the others.*

*Similarly, the user checks if the play button works correctly. The //assert? expression clearly reveals that the button state remains LOW (❌) despite having set a pull-up resistor. Verifying the hardware connections, the user spots that a jumper wire was misplaced, and after fixing it, the button asserts with ✅ in its default HIGH state.*

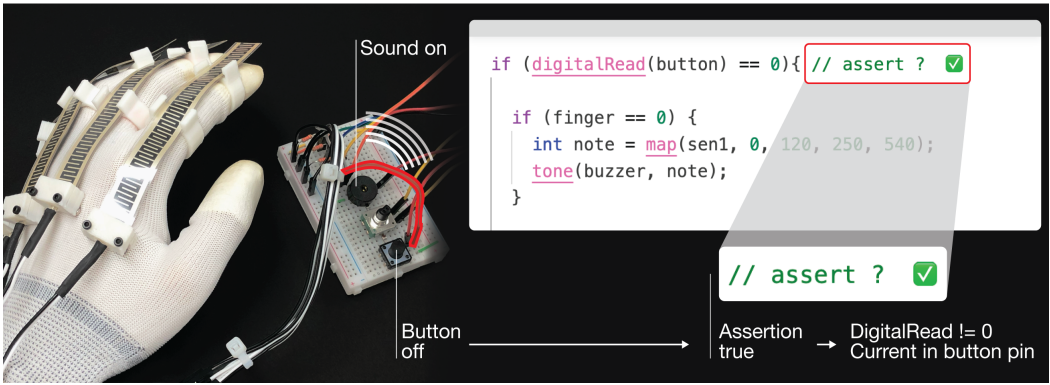

Fig. 5. Monitoring the state of the pull-up resistor button using the //assert? expression.

## 4.3 Design Goal 2: Real-time tracing the execution flow of a program with line highlight

Logs are updated in *real-time*, saving users from needing to record hardware events with a logic analyzer and examine them later in a different editor [9, 31]. With INLINE it is possible to infer which line is executed by observing which logs are updated. Furthermore, when logs are refreshed, their corresponding code lines are briefly highlighted with a yellow overlay, demarking a visible change in the state (Figure 6). In sum, the real-time behavior of the INLINE logs consists of allowing users to see a *live* [38] annotation displaying the latest updated log value triggered at a specific line of code, while simultaneously being able to observe the occurrence of the event that triggered the log by monitoring when the annotations changes.

> *In our example, the user adds a rotary encoder to the circuit for controlling the BPM of the sequencer. Incremental encoders require reading the digital status of their two out-of-phase output pins and computing the quadrature to determine the rotation direction. However, the code that handles the different quadrature cases is not working correctly. INLINE highlights the executed lines, showing that some if-statement branches are unreachable. The user then checks the hardware connections and spots a missing wire.*

### 4.4 Design Goal 3: Filter and manipulate logs with written textual expressions

A rich expression language built-in INLINE allows manipulating, altering the visualization, and filtering the hardware logs without changing the code nor requiring compiling and re-uploading the code to the Arduino. In this sense, expressions are an example of what Bret Victor defined as *explorable explanations*[7]. Expressions are specified as code comments (similar to other markup languages used to annotate comments, such as JSDoc and Javadoc), and so do not interfere with existing code and have the implicit benefit of documenting the user's intention.

The expression language includes commands for checking assertions and identities (`assert`, `is`), filtering and mapping data (`filter`, `map`), applying conditional thresholds (`above`, `below`, `between`), saving logs to variables for later comparisons (`save`, `count`, `add`, `min`, `max`), storing values in log files (`log`), changing graphical representations (`volt`, `print`) and displaying graphs (`graph`, `hist`). Table 7 shows all the expressions, their descriptions, and usages. However, it is worth noting that the expression language is extendable, as commands are plain JavaScript snippets evaluated at run time by the editor's engine. The expressions executes individually, and independently without affecting each other. The user can, however, store and share log values across code lines using variables. For example, the expression `//save varName?` create a scoped named variable *$varName* accessible from other expressions. INLINE also supports built-in variables to disambiguate among hardware logs: the default log at a line is stored in the $$ variable, while, in case of multiple statements on a single line, these are stored in a list and can be accessed by index ($1, $2, $3...).

---

[7]http://worrydream.com/ExplorableExplanations

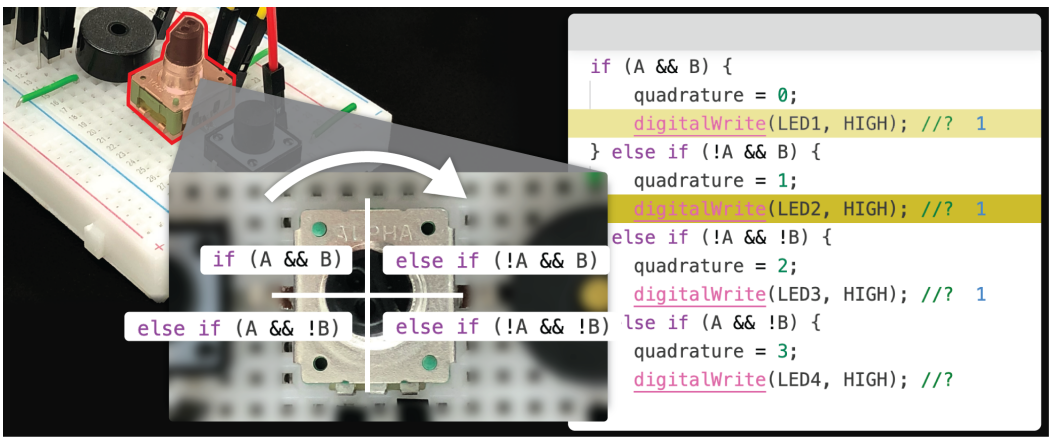

Fig. 6. The highlighting feature that shows the code execution flow helps users to spot unreached branches of the four conditions when rotating the physical encoder.





| Category | Command | Description | Example |
|---|---|---|---|
| **Identity Tests** | *assert* | returns ✅ if the log evaluates to true, otherwise ❌. | `// assert?` ✅<br>`// assert?` ❌ |
| | *is* | checks whether the log is equal to the input value or not. | `// is 1?` `true`<br>`// is 0?` `false` |
| **Thresholds** | *above* | returns the log value if greater (>) than the threshold. | `// above 100?` `508` |
| | *below* | returns the log value if smaller (<) than the threshold. | `// below 100?` `None` |
| | *between* | returns the log value if low <= input <= up. | `// between 100, 200?` `None` |
| **Storing Variables** | *save* | saves the log value in a variable. Recall the variable using the $ operator before the variable name. | `// save x?` `212` |
| | *count* | increments a counter of the log, and stores it in a variable. Recall the variable using the $ operator before the variable name. | `// count y?` `467` |
| | *add* | accumulates the log value, and stores it in a variable. Recall the variable using the $ operator. | `// add z?` `234279` |
| | *min* | stores the minimum log in a variable. | `// min a?` `0` |
| | *max* | stores the maximum log in a variable. | `// max b?` `1023` |
| **Output** | *print* | prints on the line the log and other additional values (default output). Other commands cannot be chained after this output command. | `// print $x, $y?` `, 1023, 831` |
| | *graph* | draws a line graph based on the log and other additional numerical values. Other commands cannot be chained after this output command. | `// graph $$, $x?` `510, 510, 212` |
| | |  | |
| | *hist* | draws a histogram based on the log and optional additional numerical values. Other commands cannot be chained after this output command. | `// hist?` `826` |
| | |  | |
| | *log* | logs the input value to a file (specify name or default is logs.txt). Other commands can be chained to this one. | `// log myLog.txt ?` `509` |
| | *volt* | shows the analog input as voltage (specify reference voltage or default is 5V). | `// volt?` `2.49` |
| **Advanced** | *map* | takes a log and maps it with a function to an output. | `// map x => x / 2?` `480` |
| | *filter* | takes a log and returns it if the filter function returns true. | `// filter x=> x%2 == 0?` `960` |

Fig. 7. The full list of expressions supported by Inline, divided into five categories. Each line shows the command, a description, and an example of usage.

Finally, expressions can be concatenated using the *pipe* operator '|'. Each expression takes the input value either from the hardware log at a specific line (i.e., the implicit variable $$) or from the value computed by the previous expression. For example, the expression *//above 10 | below 20 | assert?* checks whether a hardware log is between the values 10 and 20 (a shorthand for *//between 10, 20 | assert?*).





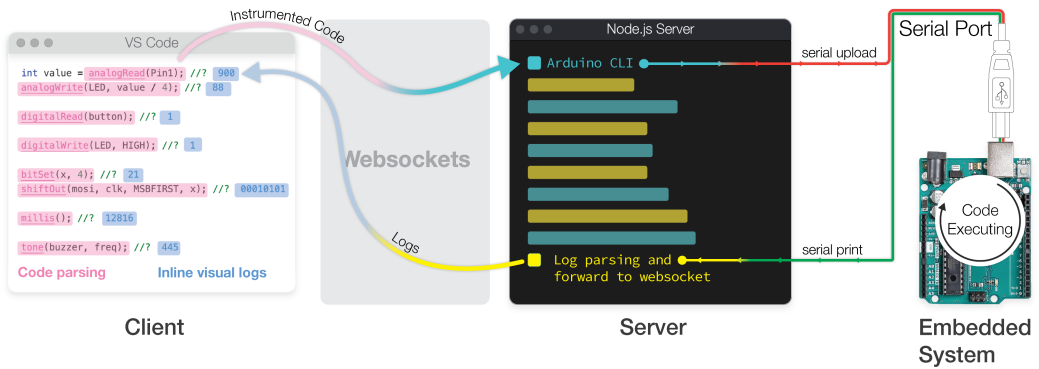

Fig. 8. The overall system architecture, its components, and how they talk to each other.

The expression *//map x => 100∗x/1024 | log logs.txt | graph $$?* is more complex and takes the log from an analog input (0-1024), maps it between 0 and 100, stores it in a time-stamped text file, and plots it as a line graph along with the original log value. Very complex expressions can also be achieved by tapping into the ecosystem of valid JavaScript code and built-in functions or libraries (e.g., Math, Date...).

> *Back to the walk-through, the user needs to map the fingers' motion to the sound (frequencies) played on the buzzer. Firstly, to determine the sensors' operation range, the user chooses the min and max expressions to track the lowest and highest output values. These expressions are concatenated in a single statement like so: //min low | max high?. Alternatively, the user can plot a histogram with //hist? and inspect the range of distribution of values. Then, to map the analog readings to frequencies, the user relies on the built-in Arduino function map (not to be confused with the INLINE expression //map?) and then checks whether its output is within the expected boundary conditions (e.g., //between 1, 800 | assert?). The output value is finally used to play a tone of a specific frequency with the function tone. Finally, the expression //? placed next to tone shows the real value in Hz of the played frequency, which might be slightly different than the numerical parameters passed to the tone function. By saving the mapped frequency value on the line of the map function (//save mappedFreq?), the user can now plot the two logs on a single graph and compare them directly (//graph $mappedFreq?).*

## 5  SYSTEM ARCHITECTURE AND IMPLEMENTATION DETAIL

Inline is implemented as an extension for the Visual Studio Code (VSCode) Insiders editor[8] and a Node.js back-end server interfacing with an AVR-based Arduino board via USB serial. Both the server[9] and the client extensions[10] were written in TypeScript. The code is open-source and available on GitHub.

The VSCode extension (client) is responsible for parsing and instrumenting the Arduino code and visualizing within the text editor the debugging logs received from the connected hardware. The user's primary way to interact with the extension is through a side panel with a graphical user interface for basic control (Figure 1). The interface includes buttons for uploading the instrumented

---

code or its release (not instrumented) version and toggling the line highlight and the inline code annotations. Finally, it also features an interactive menu with a list of expressions, their explanation, and snippets of code that can be copied with a click for quick insertion in the code. This side panel was developed using the Svelte[11] reactive framework.

The server provides an abstraction layer for the connected Arduino device and forwards real-time logs and diagnostic messages back to the extension. The server also wraps the Arduino CLI toolchain, allowing the extension to compile the Arduino sketch C/C++ code for a specific architecture and upload it onto the target board. The extension and the server communicate over websockets[12], enabling both local and remote development—i.e., the server can run locally or remotely on the same machine to which the Arduino board is connected.

## 5.1 Code parsing and instrumentation

When the user clicks on the *upload code* button, the user code is parsed, instrumented, and then sent to the server to be uploaded to the connected hardware board. Parsing of C/C++ Arduino code is achieved with a custom LALR(1) parser generated with Jison[13]. The parser checks if the code is valid, extracts the relevant function calls (e.g., function name, parameters, line of the text), and replaces them with their instrumented version. The editor also generates on-the-fly the library code that needs to be bundled up and uploaded together with the user's code on the target board.

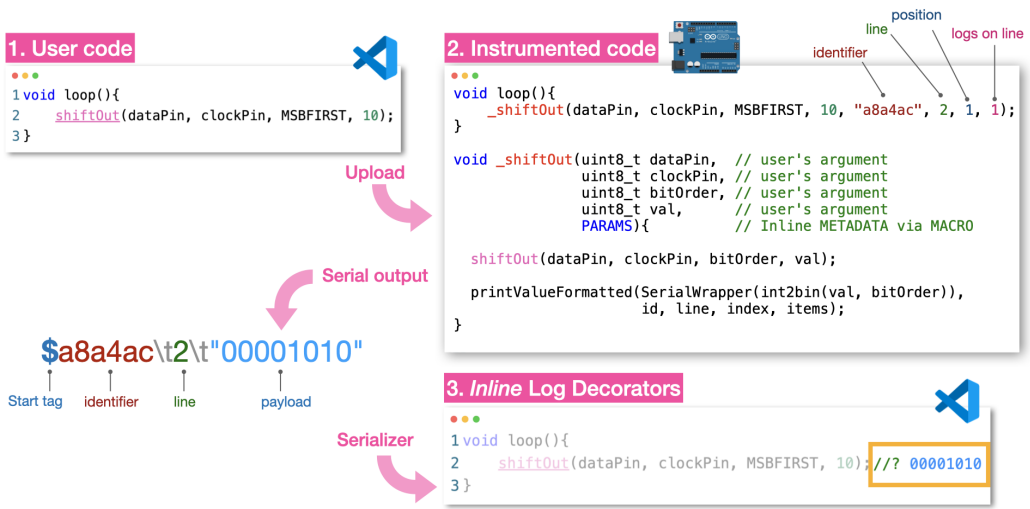

Fig. 9. An example of how code instrumentation was achieved for a single log.

The instrumented function call is simply a wrapper of the original Arduino function, augmented with metadata—e.g., a unique identifier of the function signature (i.e., a hash signature), the arguments supplied by the user, the line number where it occurs in the code, the total number of logs for the line, and, in case of multiple ones, the relative positional index. At runtime, when the code is executed on the target board, the instrumented function executes a call to the original Arduino function (the one that was wrapped), but also injects a print statement that is serially transmitted to the code editor containing the metadata discussed above. Figure 9 shows an example of this process with the Arduino function `shiftOut`. This function is used to serialize a *byte* (e.g., '10'), and the log produces its corresponding binary representation with the most significant bit (MSB) first. All Arduino functions were instrumented in a similar way, and some of the functions include more logic steps beyond logging. For example, the `analogWrite` function produces a PWM signal and internally measures the effective duty-cycle that was produced by the call, which is eventually logged.

---

[11]https://svelte.dev

[12]https://socket.io

[13]https://gerhobbelt.github.io/jison

The logs from all the instrumented functions are received by the editor via serial and are transformed into a collection of observable streams using the RxJS library[14]. Each event represents a unique observable object that updates independently from the others. When a specific event is fired, the metadata sent from the Arduino are extracted and checked to see if they match the function in the code at the same location. The two unique identifiers of the functions (the one executing on the hardware and the one in the code editor) are checked, and if there is no match, it means that the code is "dirty" and no longer maps to the log. For example, an argument might have been changed, or the line of the function call, or the order of function calls on the line. Finally, if there is a match, the code editor executes the expression at the line (if any) and decorates the active editor using the result, with the specified visual—line highlights, inline text, or graphs—as described in the next section.

## 5.2 Visual decorations of the code editor

With INLINE, the text editor is not limited to merely visualizing static text. Instead, it is an active interface component that changes depending on the incoming hardware logs. The editor provides several visualizations. The extension helps users recognize all the instrumented function calls (i.e., the ones that can emit logs) by highlighting them with an underline. These functions are all the main Arduino-specific functions listed on the reference page[15]. Custom syntax highlighting was achieved by specifying a *TextMate grammar*[16] and CSS styling.

To visualize the logs within the editor as colored text, line highlights, and graphs, we used VSCode decorators (e.g., foreground and background colors). These decorators can be refreshed with specific timing and can be customized with CSS scripting. For example, the yellow line highlight was implemented by changing the text's background color of the whole line for a fraction of a second. The decorators that appear as charts (histogram and line graphs) are implemented using the experimental *webview inset API*, which allows displaying custom HTML/JavaScript content within the editor. The graphs were implemented using the p5.js library, and these communicate with the rest of the extension via local sockets.

## 5.3 Expression language via functional programming

An important feature of INLINE is the expressions that allow manipulating and filtering the content and the visual representation of the logs during the execution of a hardware program. We developed a custom expression language using context-free grammar and generated a parser to transpile the expression code to plain JavaScript. Each expression sequence is finally executed at runtime (on save) using the VSCode built-in JavaScript engine (i.e., using the built-in `eval` function) but in its sandbox, using a *local shared context* (an Object) to prevent side effects across expressions. For example, the expression `assert` is converted to the equivalent JavaScript code `this.pipe` (`this.assert`, `this.output`('inline')( ))(`$$`), where `this` represents the context of the execution of the expression to be evaluated and `$$` is substituted at runtime with the value of the

---

hardware log received from the serial stream. To support the concatenation of *piped* expressions, we rely on functional programming by creating partially applied functions. In the example above, it is clear that the context executes a *pipe* function with two parameters—the function `assert` and the partially-applied function `output`, which specifies the format of visualization (i.e., *inline*, *graph* or *hist*) and takes no additional parameters. Note that the output function is always the last one in a sequence of expressions, and if none is specified, the expression defaults to `'inline'` presentation.

Using partially-applied functions allows concatenating expressions, each with its parameters, but always a single output. For example the sequence `//min low | max high | graph $high, $low?` is internally constructed as `this.pipe(this.min('low'), this.max('high'), this.output('linegraph') (this.high, this.low))($$)`. To explain this, the context executes a pipe function that takes three partially applied functions and a single input parameter `$$`—the actual value of the log, which is substituted to the value at run-time. The log value is passed from one function to the next one, in a purely functional programming style. Specifically, the function `min` takes the log value `$$` and stores it in a variable `$low` with the minimum value encountered so far. Then it passes the log to the next function, `max`, which does the same. In other words, functions are *curied* so that they always take a single input (from the previous function in the pipe) and output a single value (passed to the next function). Eventually, the log reaches the output function `output('linegraph')` which is partially applied with the minimum and maximum stored log (the variables `this.low` and `this.high`) and takes as input the current value of the log `$$` passed from the previous functions. The result of this entire expression is a graph displaying the three values of the minimum, maximum, and current logs.

## 6 USER STUDY

To test how users received Inline and its features, we conducted a preliminary study using five debugging exercises followed by interviews.

### 6.1 Participants

We recruited twelve makers, 8 male and 4 female, aged 20-34 years (avg=25.7, sd=4.0) among computer science, engineering, and design students from our institution. Two of the participants also participated in the formative study eight months prior. All participants reported familiarity with physical computing (avg: 4.2 of 7, sd = 1.4) and software development (avg: 4.1 of 7, sd = 1.8). They received a compensation of 20 USD in the local currency.

### 6.2 Material and Method

The study exercises involved writing code and finding software and hardware bugs in partially complete code snippets and pre-assembled breadboarded circuits. Each exercise involved finding 2-3 bugs (e.g., incorrect choice of function, missed pin initialization, miswiring, identifying ranges of proper values, etc…) for a total of 11. For details about the tasks and bugs, please refer to Table 2 and Figure 10. The tasks were created to encourage users to utilize various features of the system and language expressions. However, in order to enhance the ecological validity of the tasks, users were not artificially restricted or directed towards predefined debugging strategies, specific commands, or expressions, but were free to choose their own debugging methods. For each task, users received the electronic schematics of the breadboarded circuit they needed to debug, along with the template source code that they were expected to modify. These materials can be accessed online[17].

After collecting demographics, we briefly introduced participants to Inline with an introductory tutorial video and a 10-minute live demo of the expression language. Participants were told to

---

[17]https://github.com/makelab-kaist/inline-tasks-material

"fix" the circuits using any combination of code, INLINE features, or hardware parts. They were not told how many bugs to look for, and tasks were completed in the same order (1 through 5). This part of the study was screen-recorded for analysis and lasted up to 90 minutes. Upon completion, we conducted a 30-minute semi-structured interview centered around the task outcomes and observed usage patterns with INLINE. Interviews were transcribed, translated, and analyzed using open and axial coding methods. Three authors were involved in this process. Each researcher assigned one generic and one specific code to every quote. By grouping by code, they identified 7 themes (debugging without code modification, bridging the hardware feedback, and code execution, pinpointing errors, interaction/visualization for collaboration, contextualizing log data, expression, and simplicity/convenience). In cases where quotes had multiple codes, the raters discussed and agreed on a single code, prioritizing based on relevance. Lastly, we collected scores from participants for the System Usability Scale [25] and NASA TLX [12].

Table 2. Descriptions of the five tasks in the user study.

| Tasks | Goal | Problems and Solutions |
|---|---|---|
| 1 | Adjust LED brightness using a potentiometer | (HW/SW) Reconnect the LED to a PWM pin, not a digital one |
| | | (HW/SW) Replace digitalRead with analogRead |
| | | (HW/SW) Map analogRead values to a 0-255 range |
| 2 | Toggle the LED with a button | (HW/SW) Set button pinMode as "INPUT_PULLUP" |
| | | (SW) Use "==" for equality, not "=" |
| | | (HW/SW) Replace analogRead with digitalRead |
| 3 | Alternate two LEDs using random values | (HW/SW) Set pinMode for LEDs as "OUTPUT" |
| | | (SW) Adjust if-else condition values to 0 and 1, not 1 and 2 |
| 4 | Play a tune and silence the buzzer using a bend sensor | (HW) Ground the buzzer |
| | | (SW) Correct the minimum and maximum value ranges |
| 5 | Update LED brightness using a button and a photoresistor | (HW/SW) Modify the upper bound of the value's current range to 9 in the map function |

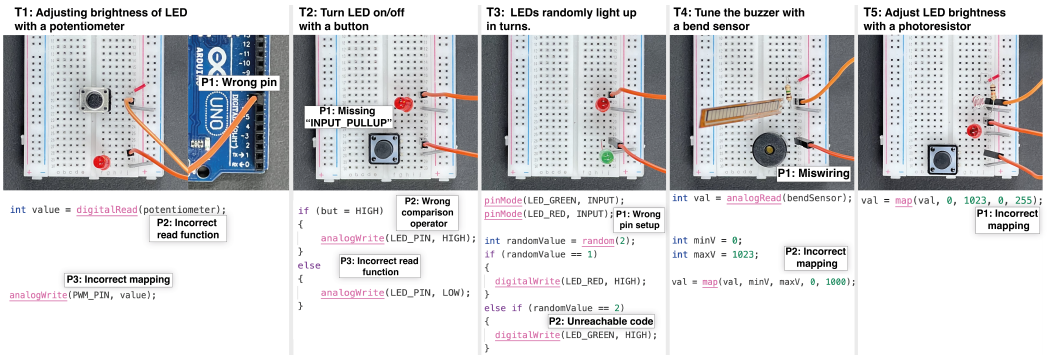

Fig. 10. Provided materials for the five tasks: assembled physical circuits and code. The white boxes annotated with answers serve as figures for explanation to the readers and were not given to the participants.

## 6.3  Results

All tasks were completed in 13 minutes (avg: 10', sd: 3'). Eight of twelve users could identify all the bugs and completed the exercises with a working code, while two participants made a single mistake. Only two users were unable to identify 3 to 4 bugs, with most of the errors in the second task (i.e., initializing internal pull-ups, using `digitalRead` instead of `analogRead`). In total, we recorded 9 events out of 132 (6%) not diagnosed as bugs.

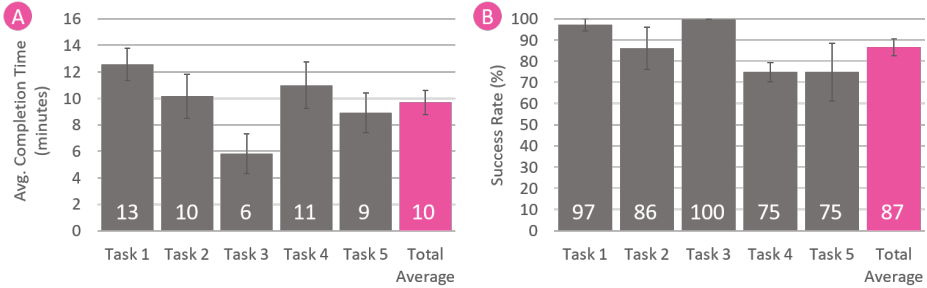

Fig. 11. Average completion time (A) and success rate (B) of all five tasks.

### 6.3.1 What INLINE features were used, usability, and perceived benefits.

All participants used inline logs (//?), line highlighting, and expressions during the study. The most popular expression was the basic //? (51 occurrences, 35% of total), followed by output expressions (n=42, 29%), identity tests (n=24, 16%), storing variables (n=16, 11%), and thresholds (n=13, 9%). The advanced expressions //map? and //filter? were not used. The overall usability of INLINE was valued with a SUS score of 79.6 of 100 (sd: 8.9) and an average TLX workload of 28.8 of 100 (SD: 18.2).

Figure 11 presents the completion time and success rate of five different tasks. On average, users completed each task in 10 minutes (min:5.8, max:12.6, SD:3 minutes), and performed with an 87% success rate (min:75%, max:100%, SD:13%). For completion time across different tasks, we observed a statistically significant difference between groups as determined by one-way ANOVA ($F_{(4,55)} = 2.535$, $p = .05$). A Bonferroni post-hoc test revealed that the time to complete the task was statistically significantly lower for task #3 (5.8 min, SD: 5 min, p = 0.03) than for task #1 (12.6 min, SD: 4 min). Additionally, for success rate across different tasks, there was a statistically significant difference between groups as determined by one-way ANOVA ($F_{(4,55)} = 3.421$, $p = .014$). A Bonferroni post-hoc test revealed that the success rate was statistically significantly higher for task #3 (100%, SD: 0 min, p = 0.019) than for task #4 (75%, SD: 15%).

Among the system features, all users welcomed the ability to see real-time logs of sensors (P1) and, more generally, logs directly next to the calling function without having to recompile and re-upload the code. They explained that "adding something to the code often leads to errors (P10)" and that "the waiting time for uploading [the code] is quite long for Arduino, but here [Inline] I didn't have to change the code every time (P1)". The highlighting of the execution line was also well received because it helped to identify "which branch was taken [or] that some part of the code was even executing [...] So if I see that everything is being executed, that is when everything's usually going well. So I like that highlighting. (P12)". Similarly, other users commented that the highlight helped them to pinpoint hardware bugs by following the execution flow (P3: "I could tell what is the cause and I think it really helped to speed up the process") or identifying miswiring (P4: "the system told me that the potentiometer is working but it [the line] wasn't blinking. I think that's how I found [the error]").

Other expressions that received positive comments were the //log? (P1: "It is a great advantage to be able to create log files with the completed prototype for user testing or other purposes"), the thresholds (P9: "I think it's really useful to have a feature where you can check whether the value is within the range of 0 to 255 [...] I feel this feature significantly reduces such tasks"), and identity expressions (P8: "the 'assert' command visually stands out the most"). Finally, graphs were used by all participants. Histograms were used for identifying operational ranges (P11), while line graphs were for visualizing real-time responses from sensors (P11: "I like that I can visually see how sensor values are changing as I move my hand"), and tuning input (P3: "the relational graph that is plotting the different values helps me to tune the sensor values"). P5 highly appreciated using multiple simultaneous graphs ("Actually for me it was great to have all the graphs types in different plots at different lines"), and this feature prompted other users (P11) to wish they could have used INLINE in past projects.

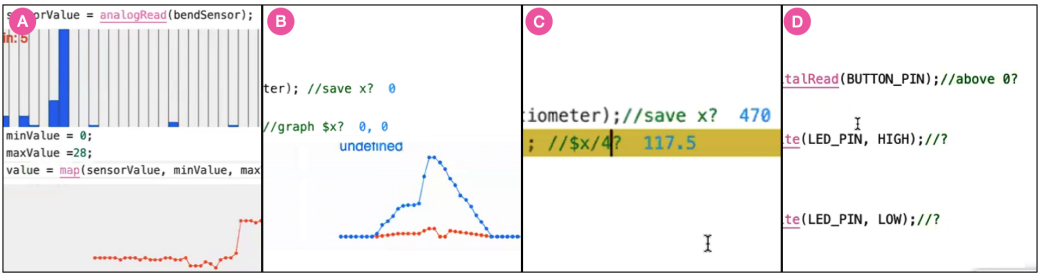

Fig. 12. Screenshots of the video recordings from the participants' monitor when solving the tasks in the user study

INLINE also received some critical comments. Some users (P10) said they did not attempt to concatenate expressions because they felt overwhelmed with options. Some users (P10, P11) felt distracted by or wanted to change the color of the line highlighter and did not know how to turn it off using the GUI panel. Other students reported multiple graphs on the same page caused them to scroll through the code to visualize the individual graphs (P9, P10). P5 suggested a better way to provide documentation of the expressions and interface buttons beyond those in the side panel, and P2 reported a laggy response of the overall interface and logs. To investigate this issue we inspected the video recording of the participants interacting with the systems and observed that the issue with the delayed response was related to uploading the code on the Arduino, a common problem faced by developers when using some of the AVR boards[18], and probably rooted to our choice of hardware (a board with limited baud rate) and the underlying Arduino-CLI tool. Further investigations are needed for conclusive reporting about the responsiveness of the system.

*6.3.2   How INLINE features were used.* The video analysis of how participants used INLINE revealed different approaches toward debugging. Several participants (P2, P4, P5, P9) initially started with the basic //? expression and later modified it with more specific intent (e.g., //volt?) or graphed the results. This specialization of intent was a common pattern, where users eyeballed the result and then tried to understand it with more specific expressions. Another typical pattern was a comparison of data either via multiple graphs (Figure 12A) or by storing a variable and then using it in a graph or in another expression (Figure 12B-C). Similarly, the //assert? expression was used as a way to trace whether specific lines were executed at least once (P9 said: "It helps you to reduce the amount of thinking a little bit when it is presented in a refined way. If it is displayed as numbers, you have to check and verify them again for instance whether they fall within the range, but if you just check the mark, you can immediately see if it has entered or not."). Another user repurposed the thresholds (e.g., //above 0?) to guard against nullish values (Figure 12D).

---

[18]http://tinyurl.com/4j65762u

Another common pattern, seen in all users, was to leave Inline expressions in the code. Although there are many reasons why participants might have left them there, P11 offered a possible explanation: "They are like annotations; being already in place helped me understand the code overall. Sometimes, we split up the work when we develop a program together. That means I might end up debugging someone else's code. When multiple people work on code, adding annotations can help others understand."

One last interesting finding was that, in some cases, the users could identify a bug but simply did not know how to fix it. For example, P4 explained that "When I saw the data from the potentiometer [using Inline ], I could easily know that the value was not appropriately passed to the LED, but I could not figure out how to modify it." Similarly, P7 made a similar statement "I didn't know how to fix it, but intuitively, I knew the values should come out somewhat consistently." This suggests that Inline could help beginners, who do not have proper training or sufficient knowledge of embedded programming, to at least be able to identify where a problem happens, and then seek help elsewhere with an instructor or using online references.

## 7 DISCUSSION

The formative study presented earlier in this paper (section 3) highlighted three key challenges among makers for hardware-software integrated debugging. From these, we derived three **D**esign **G**oals: **DG1**) inline visualization of debugging logs; **DG2**) real-time tracking of the program's execution flow; **DG3**) filters and manipulations of logs via expressions. In our usability study, all participants adopted these three debugging strategies and provided insights into what features were mostly welcomed and recurring use patterns. We identify an increased specialization of intent when debugging, using checkpoints in the form of asserts throughout the code, intentionally leaving annotations as future documentation, and being able to identify potential errors despite perhaps being unable to solve them. While acknowledging the usefulness of various features, our participants expressed being overwhelmed with the number of features and options available within Inline, which were only briefly mentioned during the tutorial session before the debugging tasks.

From our collective findings we see opportunities for practical usage of Inline in class settings and among makers, complementing systems like ElectroTutor [49] for guided tutorials, Bifröst, ToastBoard and CircuitSense [9, 31, 53] for hardware inspection, and Strasnick's [45] work about coupling hardware with software simulations. However, comparing Inline with prior work that dealt with debugging and visualization of hardware, we stress the uniqueness of our approach in three key aspects: 1) Inline offers a *software-only approach* for visualizing hardware logs and does not require physical instrumentation or additional sensing hardware, differently from systems like ToastBoard and Pinpoint [9, 46]. Furthermore, 2) Logs are *generated, visualized, and manipulated in real-time and directly within the user's code*, without needing terminal consoles or separate graphical interfaces like in Bifröst [31]. Finally, 3) Inline works with the user's arbitrary code, and it is not limited to testing snippets or small subsets of code, like in [31, 43, 49]. These three aspects share common ground and motivations with many prior tools but also embody the main differences with past works and highlight how Inline could benefit these systems as well.

Like previous work in the domain of exploratory and live programming tools [38] we are interested in creating a system that leverages the *liveness* of elements within the code to increase the comprehension of a program and to boost the creativity and explorations of a developer. Differently from past work, we achieve this using in-situ visualizations of hardware logs that can be tuned with an expression language, rather than relying on representations on separate GUI elements [31] that could result in a loss of context. Although we did not explore alternative ways to visualize the hardware logs within the code, literature provides numerous examples, such as timelines that allow semantic and temporal grouping of logs [34], visualizations via color coding and indentations [14], or even user-customized code representations and projections [10].

## 8 LIMITATIONS AND FUTURE WORK

In this section we acknowledge some of the limitations of the the INLINE system implementation and of its evaluation, and highlight some opportunities for future research. Starting from the technical limitations, the INLINE's built-in C++ code parser is limited to a subset of the C++ language (i.e., the Arduino built-in functions) and cannot handle nested function calls. The parser also works only with the currently active editor window, meaning that it operates on a single file at a time—the one opened and in focus. A different implementation that would solve this problem could use an incremental parser such as Tree-sitter[19], combined with an extension for Arduino language[20]. By keeping track of the full AST (Abstract Syntax Tree) and by performing queries and updates on the syntax tree when needed, the system not only would perform more efficiently but also could overcome the current technical parsing limitations. Finally, INLINE offers opportunities for extension, such as tracking variables and not only built-in function calls, or adopting different frameworks and programming languages (e.g. MicroPython or JavaScript). Another technical limitation is that, by considering only the set of Arduino built-in functions, using the //? expression on a not instrumented line would result in a silent failure. This limitation could be addressed with better warnings triggered when expressions are used on lines without instrumentation. Furthermore, the logs in the current prototype have no notion of "time and order", as the logs are displayed as soon as they are received by the client. An alternative would be to construct interactive timeline visualizations, like in [34], where for instance logs triggered on the same line of code, but at a different time, could be graphically overlapped to support a comparison of distinct behaviors over time.

One last technical limitation of our prototype is the delays (approximately 5 ms) introduced by using a serial port for real-time transfer of logs and metadata. Using protocol buffers instead of ASCII text could help speed up the communication, but for time-sensitive domains, debugging boards that support higher baud rates or JTAG should be considered instead. All these technical limitations are important but should be considered as implementation trade-offs for building a working prototype. They are addressable with optimizations beyond the scope of this paper and they did not limit our explorations of the design space, nor are misaligned with the typical expectations of a research through design method [50].

Finally, we acknowledge some limitations of the user study. Participants in both studies were recruited via bulletin posting and word-of-mouth from a single institution, and future work should attempt to consider participants with various backgrounds and origins. Furthermore, most interviews reported in the paper were conducted in a language other than English, and translation might have glossed over nuances and subtle expressions. Nonetheless, we are confident that our samples are representative of the target users for which this work was originally intended. Future work will attempt to test INLINE with a broader audience by both introducing it in university courses and releasing the VSCode extension to the public. We hope to gather feedback from the maker community about the usage of the extension over time.

## 9 CONCLUSIONS

In conclusion, we presented INLINE, a programming tool that offers a simple *window* for understanding "what is happening inside the hardware of an embedded system" when a program is executing. It allows to visualize and manipulate *real-time live debugging logs* generated in the hardware directly within the text of the source code of a running program. These logs can be presented as either text or graphics, manipulated with expressions, and offer immediate insights into which lines of the code are executed and how different instructions affect the internal state of the hardware. The system was designed based on feedback from a formative study and was finally tested with 12 users to complete five debugging tasks. The results show that all users could adopt, in a very short time, all the main features of Inline and familiarize themselves with most of the expression language. Future work will delve toward better integrating our software-only approach with existing hardware probing debugging tools and expand the range of interactions to manipulate the output logs.

---

[19]https://tree-sitter.github.io
[20]https://github.com/ObserverOfTime/tree-sitter-arduino

## ACKNOWLEDGMENTS

This work was supported by the National Research Foundation of Korea (NRF) grant funded by the Korea government (MSIT) (No. 2018R1A5A7025409). Yoonji Kim was supported by the Chung-Ang University Research Grants in 2021. The authors thank Steve Hodges and Bjoern Hartmann for their early feedback about the system presented in this paper.